\documentclass{ws-procs10x7}
\usepackage{balance}

\makeindex
\begin{document}

\title{Theory of Macroscopic Quantum Dynamics in High-$T_c$ Josephson Junctions}

\author{SHIRO KAWABATA}

\address{Nanotechnology Research Institute (NRI), National Institute of 
Advanced Industrial Science and Technology (AIST), Tsukuba, 
Ibaraki, 305-8568, Japan and \\
CREST, Japan Science and Technology Corporation (JST), Kawaguchi, Saitama 332-0012, Japan
\\E-mail: s-kawabata@aist.go.jp}


\twocolumn[\maketitle\abstract{We have theoretically investigated macroscopic quantum tunneling (MQT) and the influence of nodal quasiparticles and zero energy bound states (ZES)
on MQT in  $s$-wave/ $d$-wave hybrid Josephson junctions.
In contrast to $d$-wave/$d$-wave junctions, the low-energy quasiparticle dissipation resulting from nodal quasiparticles and ZES is suppressed due to a quasiparticle-tunneling blockade effect in an isotropic $s$-wave superconductor. 
Therefore, the inherent dissipation in these junctions is found to be weak. 
This result suggests high potential of $s$-wave/$d$-wave hybrid junctions for applications in quantum information devices.
}
]

\section{Introduction}

Since the experimental observations of macroscopic quantum tunneling (MQT) in YBCO grain boundary\cite{rf:Bauch1,rf:Bauch2} and Bi2212 intrinsic\cite{rf:Inomata,rf:Jin,rf:KashiwayaMQT1,rf:KashiwayaMQT2} Josephson junctions, the macroscopic quantum dynamics of high-$T_c$ $d$-wave junctions has become a hot topic in the field of superconductor quantum electronics and quantum computation.  
Recently, the effect of low-energy quasiparticles, e.g., the nodal-quasiparticles and the zero energy Andreev bound states (ZES)\cite{rf:Kashiwaya,rf:Lofwander} on MQT in $d$-wave  junctions have been theoretically investigated.\cite{rf:Kawabata1,rf:Yokoyama,rf:Kawabata2}
It was found that, in $c$-axis type junctions, the suppression of MQT resulting from the nodal-quasiparticles is very small.\cite{rf:Kawabata1,rf:Yokoyama} 
This result is consistent with recent experimental observations.\cite{rf:Inomata,rf:Jin,rf:KashiwayaMQT1,rf:KashiwayaMQT2} 
In the case of in-plane type $d$-wave junctions, however, the ZES give a strong dissipative effect.\cite{rf:Kawabata2} 
Therefore, it is important to avoid the formation of ZES in order to observe MQT with a high quantum-to-classical crossover temperature $T^*$.

On the other hand, recently, quiet qubits consisting of a superconducting loop with an $s$-wave/$d$-wave ($s/d$) hybrid junction have been proposed.\cite{rf:Ioffe}
In the quiet qubits, a quantum two level system is spontaneously generated and therefore it is expected to be robust to decoherence caused by fluctuations of the external magnetic field.\cite{rf:Ioffe,rf:Kawabata5}
However, influence of the low-energy quasiparticle dissipation due to nodal-quasiparticles and ZES on $s/d$ quiet qubits is not yet understood.
Therefore, it is important to investigate such intrinsic dissipation effects on the macroscopic quantum dynamics in order to realize quiet qubits.

 In this paper, motivated by the above studies, we have investigated investigate MQT in $s/d$ hybrid Josephson junctions [see Fig. 1(a)].\cite{rf:Kawabata6}
  In contrast with $d/d$ junctions,\cite{rf:Kawabata1,rf:Kawabata2} we have shown that the influence of the low-energy quasiparticle dissipation is suppressed by a small but finite isotropic gap in the $s$-wave superconductor. 
 Therefore, weak quasiparticle dissipation is expected in such junctions.

\section{Effective action}
In the following we derive the effective action for $s$/$d$ junctions with a clean insulating barrier and without extrinsic dissipation, e.g., an ohmic dissipation in a shunt resistance.
The partition function of a junction can be described by a functional integral over the macroscopic variable (the phase difference $\phi$),\cite{rf:Eckern}  i.e., 
$
{\cal  Z} 
= 
\int 
{\cal D} \phi (\tau) 
\exp
\left(
  - {\cal S}_{\mathrm{eff}}[\phi]/\hbar
\right)
.
$
%
%
%
%
\begin{figure}[t]
\begin{center}
\includegraphics[width=7.0cm]{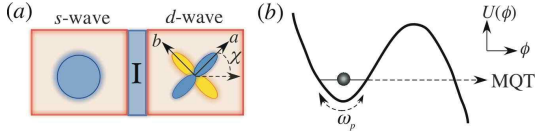}
\caption{(a) Schematic of an in-plane $s$-wave/$d$-wave hybrid Josephson junction. $a$ and $b$ denote the crystalline axes of the $d$-wave superconductor, and $\chi$ is the mismatch angle between the normal of the insulating barrier (I) and the $a$-axis.
(b) Potential $U(\phi)$ v.s. the phase difference  $\phi$ between two superconductors and 
$\omega_p$ is the Josephson plasma frequency of the junction.
}
\end{center}
\end{figure}
We consider the high barrier limit, where  $z_0\equiv m w_0/\hbar^2 k_F  \gg 1$.
Here $m$ is the electron mass, $w_0$ is the height of the delta function-type insulating-barrier I (see Fig. 1(a)), and $k_F$ is the Fermi wave length.
In this limit, the effective action ${\cal S}_{\mathrm{eff}}$ is given by ${\cal S}_{\mathrm{eff}}={\cal S}_{\mathrm{0}}+{\cal S}_\alpha$,
\begin{eqnarray}
{\cal S}_{\mathrm{0}}[\phi]
&= &
\int_{0}^{\hbar \beta} d \tau 
\left[
   \frac{M}{2} 
   \left(
   \frac{\partial \phi ( \tau) }{\partial \tau}
   \right)^2
   + 
   U(\phi)
\right],
\nonumber\\
{\cal S}_\alpha[\phi]
&=&
-
\int_{0}^{\hbar \beta} d \tau d \tau'
  \alpha (\tau - \tau') 
  \nonumber\\
  &\times&\cos \frac{\phi(\tau) - \phi (\tau') }{2}
.
  \label{eqn:alpha2}
\end{eqnarray}
In this equation, $\beta = 1 /k_B T$, $M
 = 
 C \left( \hbar/2 e\right)^2
$
is the mass ($C$ is the capacitance of the junction) and the potential $U(\phi)$ can be described by 
\begin{eqnarray}
 U(\phi) 
 = 
\frac{\hbar }{2e} 
\left[
    \int_0^1 d \lambda \ \phi  I_J (\lambda \phi) - \phi \  I_{ext}
\right],
\end{eqnarray}
where $I_J$ is the Josephson current and $ I_{ext}$ is the external bias current.
The dissipation kernel $\alpha(\tau)$ is related to the quasiparticle current $I_{\mathrm{qp}} (V)$
under the constant bias voltage $V$ by
$
\alpha(\tau) 
=
(\hbar / e)
\int_0^\infty d \omega / (2 \pi ) \exp \left(  -\omega \tau \right)
  $ $  \times I_{\mathrm{qp}} \left( V= \hbar \omega / e \right)
$
at zero temperature.

In order to investigate effects of the nodal-quasiparticles and the ZES on MQT, we derive effective actions for two types of $s$/$d_\chi$ junctions, i.e., $\chi=0$ and $\pi/4$, with $\chi$ being the mismatch angle between the normal of the insulating barrier (I) and the crystalline axis of the $d$-wave superconductor (see Fig. 1(a)).
In the case of $s/d_0$ junctions, ZES are completely absent.\cite{rf:Kashiwaya}
On the other hand, in $s/d_{\pi/4}$ junctions, ZES are formed near the interface between I and the $d$-wave superconductor $d_{\pi/4}$.

By solving the Bogoliubov-de Gennes equation,\cite{rf:Kashiwaya} we obtain analytical expressions for the potential as
\begin{eqnarray}
U(\phi) 
\! \! 
\approx 
\! \! 
\left\{
\begin{array}{cl}
\displaystyle{- \frac{\hbar I_{C_1}}{2e}\left(  \cos \phi + \eta \phi \right) }
&    
\mbox{for} \    \mbox{$s/d_0$}
 \\
 \\
\displaystyle{- \frac{\hbar I_{C_2}}{2e}\left( -  \frac{\cos   2 \phi}{2}  +  \eta \phi  \right)}    
&    
\mbox{for} \   \mbox{$s/d_{\pi/4}$}
\end{array}
\right.
\nonumber\\
\end{eqnarray}
where $I_{C_1} (I_{C_2})$ is the Josephson critical current for the $s/d_{0} $ $(s/d_{\pi/4})$ junction, and $\eta \equiv I_{\mathrm{ext}}/I_{C_1(C_2)}$.
The dissipation kernel $\alpha(\tau)$ is given by
\begin{eqnarray}
\alpha(\tau)
\! \! 
\approx 
\! \! 
\left\{
\begin{array}{cl}
\displaystyle{
  \frac{3 \hbar}{8 \sqrt{2} \pi}
\frac{ R_Q }{ R_N \varepsilon}
\frac{1}{\tau^2}
K_1 \left(  \frac{\Delta_s |\tau|}{\hbar} \right)
   }
& 
 \mbox{for}\  \mbox{$s/d_{0}$}
 \\
 \\
\displaystyle{
  \frac{6}{5 \hbar}
\frac{ R_Q \Delta_s \Delta_d}{ R_N }
K_1 \left(  \frac{\Delta_s |\tau|}{\hbar} \right)   }
& 
 \mbox{for}\   \mbox{$s/d_{\pi/4}$}\\
\end{array}
\right.
\nonumber\\
\end{eqnarray}
where $R_Q$ is the resistance quantum, $R_N$ is the normal state resistance, $\Delta_{s(d)}$ is the superconducting gap for $s$($d$)-wave superconductors, $\varepsilon=\Delta_d/\Delta_s$, and $K_1$ is the modified Bessel function.
For $ |\tau| \gg \hbar / \Delta_s$ the dissipation kernel decays exponentially as a function of the imaginary time $\tau$, i.e., 
\begin{eqnarray}
\alpha(\tau)
\! \! 
\approx 
\! \! 
\left\{
\begin{array}{c}
\displaystyle{
  \frac{3 \hbar^{3/2}}{16 \sqrt{\pi}}
\frac{ R_Q  \sqrt{\Delta_s}}{ R_N \Delta_d}
\frac{1}{\left| \tau \right|^{5/2}}
e^{  - \frac{\Delta_s |\tau|}{\hbar} }
   }
\\
\displaystyle{\quad \quad \quad   \quad \quad \quad   \mbox{for} \quad \mbox{$s/d_{0}$}}
\\
\displaystyle{
  \frac{6 \sqrt{\pi} }{5 \sqrt{2 \hbar}}
\frac{ R_Q \sqrt{\Delta_s} \Delta_d}{ R_N }
\frac{1}{\sqrt{\left| \tau \right|}}
e^{  - \frac{\Delta_s |\tau|}{\hbar} }
}
\\ 
\displaystyle{\quad \quad \quad   \quad \quad \quad \quad  \mbox{for}\   \mbox{$s/d_{\pi/4}$}}
\\
\end{array}
\right.
.
\end{eqnarray}
If the phase varies slowly with the time scale given by $\hbar /\Delta_s$, then we can expand $\phi(\tau) - \phi (\tau')$ in Eq. (\ref{eqn:alpha2}) about $\tau=\tau'$.
This gives 
\begin{eqnarray}
S_\alpha[\phi]
\approx
\frac{\delta C}{2}
\int_{0}^{\hbar \beta} d \tau 
   \left[
   \frac{\hbar}{2e}
   \frac{   \partial \phi ( \tau) }{\partial \tau}
   \right]^2
   .
\end{eqnarray}
Hence, the dissipation action ${\cal S}_\alpha$ acts as a kinetic term so that the effect of the quasiparticles results in an increase of the capacitance, $C \to C + \delta C \equiv C_{ren}$.
This indicates that the quasiparticle dissipation in $s$/$d$ junctions is qualitatively weaker than that in in-plane $d/d$-wave junctions in which the super-ohmic ($\alpha(\tau) \sim |\tau|^{-3}$)\cite{rf:Kawabata1,rf:Yokoyama} or ohmic dissipation ($\alpha(\tau) \sim \tau^{-2}$)\cite{rf:Kawabata2} appears.

\section{MQT}
Next, we will investigate MQT in $s$/$d$ junctions
The MQT escape rate from the metastable potential (Fig. 1(b)) at zero temperature is given by\cite{rf:MQT1}
\begin{eqnarray}
\Gamma
=
\lim_{\beta \to \infty} \frac{2}{\beta} \mbox{ Im}\ln {\cal Z}.
\end{eqnarray}
By using the Caldeira and Leggett theory,\cite{rf:Caldeira} the MQT rate is approximated as 
\begin{eqnarray}
\Gamma(\eta)
=
\frac{\omega_p(\eta)}{2 \pi}\sqrt{120 \pi B(\eta) }\  \exp [ -B(\eta)]
,
\end{eqnarray}
where 
\begin{eqnarray}
\omega_p (\eta)=  \sqrt{\frac{\hbar a_i I_{C_i} }{2 e M_\mathrm{ren}}}(1-\eta^2)^{\frac{1}{4}},
\end{eqnarray}
is the Josephson plasma frequency ($i=1$ for $d_0$, $i=2$ for $d_{\pi/4}$, $a_1=1$, $a_2=2$, and $M_\mathrm{ren}=(\hbar/2 e)^2 C_\mathrm{ren}$) and $B(\eta)= {\cal S}_{\mathrm{eff}}[\phi_B]/\hbar$ is the bounce exponent, which is the value of the the action ${\cal S}_{\mathrm{eff}}$ evaluated along the bounce trajectory $\phi_B(\tau)$.
The analytic expression for the bounce exponent is given by
\begin{eqnarray}
B(\eta)=\frac{b_i}{e}
  \sqrt{ \frac{2 e}{\hbar}I_{C_i} M_\mathrm{ren}}
  \left(
  1 -
 \eta^2
\right)^{\frac{5}{4}}
,
\end{eqnarray}
where $b_1=12/5$ and $b_2=3\sqrt{2}/5$.
In actual MQT experiments, the switching current distribution $P(\eta)$ is measured, where $P(\eta)$ is related to the MQT rate $\Gamma(\eta)$ as 
\begin{eqnarray}
 P(\eta)=\frac{1}{v}
\Gamma(\eta) \exp
\left[
-\frac{1}{v}
\int_0^{\eta} \Gamma(\eta') d \eta'
\right]
.
\end{eqnarray}
Here $v \equiv \left| d \eta / d t \right| $ is the sweep rate of the external bias current.
At high temperatures, the thermally activated (TA) decay dominates the escape process.
Then the escape rate is given by the Kramers formula\cite{rf:MQT1}
\begin{eqnarray}
\Gamma=\frac{\omega_p}{2 \pi }\exp \left( -\frac{ U_0}{ k_B T}\right),
\end{eqnarray}
where $U_0$ is the barrier height.
Below the crossover temperature $T^*$, the escape process is dominated by MQT. 
Note that $T^*$ is reduced in the presence of dissipation.

In order to see explicitly the effect of the quasiparticle dissipation on MQT, we numerically estimated $T^*$.
We determined  $T^*$ from the relation 
\begin{eqnarray}
\sigma^\mathrm{TA} (T^*) = \sigma^\mathrm{MQT},
\end{eqnarray}
where $\sigma^\mathrm{TA} (T)$ and $\sigma^\mathrm{MQT} $ are the standard deviation of $P(\eta)$ for the temperature-dependent TA process and  the temperature-independent MQT process, respectively.
Presently, no experimental data are available for ideal highly under-damped $s/d$ junctions with large McCumber parameters $\beta_M = (2e/\hbar) I_C C R_{\mathrm{sg}}^2 \gg 1$ ($R_{\mathrm{sg}}$ is the subgap resistance).
Thus, we estimate $T^*$ by using the parameters for an actual Nb/Au/YBCO junction\cite{rf:Smilde} ($C=0.60$ pF, $ I_C= 95.2$ $\mu$A at $\chi=0$, and $R_N=3.68$ $\Omega$) in which $\beta_M \approx 1.5$.
We also assume $ I_{C_1}=I_{C_2}=95.2$ $\mu$A for simplicity, $\Delta_s=\Delta_\mathrm{Nb}=1.55$ meV, $\Delta_d=\Delta_\mathrm{YBCO}=20.0$ meV, and $ v  I_{C_i} = 0.0424$ A/s.
In the case of $s/d_0$ junctions, we obtain $T^* = 336 $mK for the dissipationless case ($C_\mathrm{ren}= C$) and $T^* = 333 $mK for the dissipative case ($C_\mathrm{ren}= C + \delta C$).
Thus, the influence of the nodal-quasiparticle is negligibly small.
On the other hand, in the case of $s/d_{\pi/4}$ junctions, we obtain $T^* = 601 $mK for the dissipationless case and $T^* = 101 $mK for the dissipative case.
As expected, the ZES have a larger influence on MQT than the nodal-quasiparticles.
However, the $T^*$ suppression is small enough to allow experimental observations of MQT.

\section{Summary}

In conclusion, MQT in the  $s$/$d$ hybrid Josephson junctions with the perfect insulating barrier has been theoretically investigated using the path integral method.
The effect of the low energy quasiparticles on MQT is found to be weak.
 This can be attributed to the quasiparticle-tunneling blockade effect in the $s$-wave superconductor.
 We also investigated MQT in a realistic $s$/$d$ junction and showed that the expected $T^*$ is relatively high in spite of the small $\beta_M$.
   These results strongly indicate that $s$/$d$ hybrid junctions are expected to be highly applicable to quantum computers.

\section*{Acknowledgments}
I would like to thank my co-workers A.~A.~Golubov, Ariando, C.~J.~M.~Verwijs, H.~Hilgenkamp, and J.~R.~Kirtley. 
 
\balance

\end{document}